# PLASMA LENS FOR THE FOCUSING OF POSITRON BUNCHES

*D. S. Bondar[1], C. A. Lindstrøm[2], V. I. Maslov[1,3], and I. N. Onishchenko[1]*

[1]*National Science Center "Kharkiv Institute of Physics and Technology", Kharkiv, Ukraine*
[2]*Department of Physics, University of Oslo, Oslo, Norway*
[3]*Deutsches Elektronen-Synchrotron DESY, Hamburg, Germany*
E-mail: bondar.ds@yahoo.com

The development of effective focusing schemes for positron bunches in plasma accelerators remains a significant challenge, as nonlinear regimes fail to create stable focusing channels for positrons. This work presents a method for focusing and improving the quality of positron bunches using a plasma lens operating in the linear regime. Through numerical simulations, we investigate two distinct focused positron bunch profiles: a purely Gaussian bunch and an elongated, flat-top bunch with Gaussian rising and falling edges. For both configurations, the results demonstrate the capability to achieve high-quality transverse focusing. Furthermore, beyond focusing, the proposed system enables potential possibility to reduce energy spread of positron bunches of the sequence after precursor.

## INTRODUCTION

The field of plasma-based acceleration has grown considerably since the early 2000s, motivated by the prospect of achieving accelerating gradients far exceeding those obtainable with conventional radiofrequency accelerators. High-efficiency acceleration of electron bunches in the plasma wakefield was studied experimentally and by numerical simulation in particular in [1–10].

The basic idea of compensating the space charge of an electron bunch with plasma evolved into the concept of focusing the bunch with its own wakefield [2]. Another approach that follows from this is the use of a precursor bunch to create an focusing wakefield, which allows uniform focusing for next bunches of the sequence. In this paper, this method is successfully applied to positron bunches.

Although much progress has been made in electron acceleration using plasma wakefield, positron acceleration remains more challenging because of the inherently different interaction of positrons with the plasma medium. In the ideal "blowout" or "bubble" regime, an intense electron bunch expels plasma electrons to create a nearly ion–filled cavity that provides linear focusing for trailing electrons, yet the focusing possibilities for positrons are very limited because of the ion background.

These differences introduce strong challenges in saving bunch emittance and quality for positrons in plasma wakefield accelerators [11, 12].

The linear plasma wakefield regime is particularly attractive for positron acceleration because it does not lead to complete blowing out of electrons, which destabilizes the positron bunch. Particle-in-cell (PIC) simulations have shown that a hollow electron bunch driver can generate uniform accelerating fields and linear focusing forces over a region with scales comparable to typical experimental positron bunch dimensions [12, 13] These simulations indicate accelerating gradients above 5 GV/m and transverse focusing with field gradients that support low emittance growth, albeit the parameter window for robust positron transport is comparatively narrow [13, 14].

Experimental observations have also contributed to the understanding of positron focusing in plasma. The authors of [15] demonstrated plasma focusing of a 28.5 GeV positron bunch wherein the plasma lens provided simultaneous focusing in both transverse dimensions. In [16] the authors of current paper previously showed that a positron bunch in a plasma is focused worse than an electron bunch.

Many papers are also devoted to focusing electron bunches [17-25].

A number of theoretical studies have explored the fundamental differences between electron and positron interactions with plasma wakefield. It was shown [25] that in the linear regime, by careful tailoring of the drive bunch parameters and plasma density, it is possible to achieve conditions where the transverse focusing force is approximately linear with radius.

In dielectric laser accelerators (DLA), the plasma is introduced into the transport channel of a dielectric structure, where the combined effect of the plasma and dielectric walls produces high focusing fields for positrons, while simultaneously mitigating bunch breakup instabilities [26, 27]. The study [28] shows that a tapered active plasma lens can improve positron capture by 50–100% over the ILC's quarter wave transformer, with yield stability within ±1.5% for ±10% parameter variations. The study [29] explores the use of a plasma lens to enhance positron luminosity in future linear colliders, showing a potential sevenfold increase, but also highlights significant background levels that may challenge detector performance and require further mitigation. On the computational front, the use of advanced particle-in-cell (PIC) simulations has been instrumental in clarifying the dynamics of positron bunch focusing in plasma [30, 31].

In [32] it was shown that laser-driven wakefield can enable hollow electron injection and strong positron acceleration, with focusing forces over ten times stronger than in spherical bubbles and achievable under current experimental conditions.

In [33] authors show that a high-energy positron bunch undergoes nonlinear plasma focusing, leading to bunch size reduction and emittance equalization, with halo formation at higher plasma densities. The study [34] shows that thin plasma lenses can provide high focusing force and boost collider luminosity by an order of magnitude, with distinct advantages for underdense lenses in focusing high-energy electron and positron bunches.

Thus, the problem of focusing positron bunches is relevant. In this article, the authors propose a solution by positioning the bunches after bunch-precursor in the region where the head of the bunch is decelerated, the tail of the bunch is accelerated and the central part (middle) is in the zero field. Due to this, it is possible to compensate (reduce) for the energy spread and ensure focusing of the bunch.

## STATEMENT OF THE PROBLEM

The study was completed by numerical simulation. A 2d3v system with cylindrical symmetry was considered. The quantities in the text and graphs are given in normalized units. All lengths are normalized to $c/\omega_{pe}$=16.82 μm, all times are normalized to $\omega_{pe}^{-1}$=56 fs, field (electric, magnetic and force) units are normalized to $EE_0^{-1}=m_e c \omega_{pe} e^{-1}$. The electron density is $n_0=10^{17}$ cm$^{-3}$. Energy and momentum are normalized to $m_e c^2$ and $m_e c$. The bunch current is normalized to $I \cdot I_0^{-1} = m_e c^3 e^{-1} \approx 17$ kA. The magnetohydrodynamic plasma model was used. Mobile ions and cold plasma ($T_i$=0) in the absence of an external magnetic field were considered. The length of the window was $\xi_{max}$=33 $c/\omega_{pe}$ (for the case of a long bunch 50 $c/\omega_{pe}$, but this does not affect anything), the diameter (transverse dimension) of the window was $r_{max}$=5 $c/\omega_{pe}$. The initial γ factor of the bunches was γ=5. The bunch radius was 0.1 $c/\omega_{pe}$.

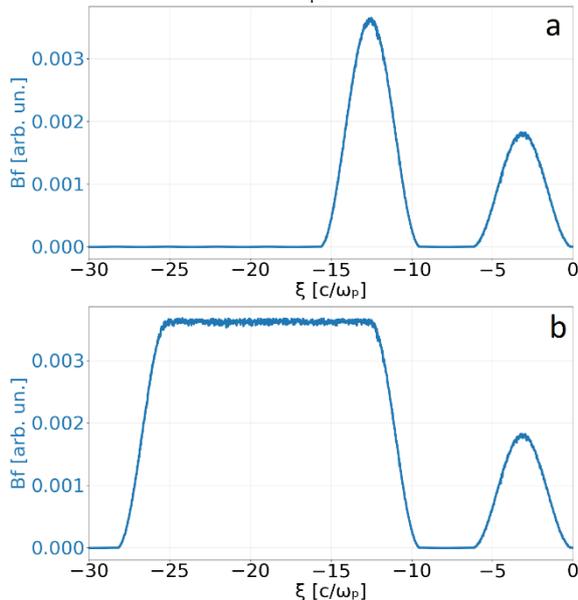

*Fig. 1. Azimuthal component $B_f$ of the magnetic field depending on the longitudinal coordinate of the system. The magnetic field is formed by the bunch current and indicates its position. (a) Short Gaussian (cosine) bunch, (b) Long bunch with homogeneous center and Gaussian head and tail.*

Two main cases were considered (Fig. 1). After bunch-precursor we studied dynamic of the positron bunch: 1) Gaussian (cosine) bunch, 2) bunch that homogeneous in the center with Gaussian head and tail.

In both cases, also sequence of several bunches after positron bunch-precursor can also be considered. The radius of the bunch is defined as the half-width of the Gaussian distribution, limited by 3σ. The current of the first bunch ($I_{b1}$=5.1 A) was chosen to be half that of the second bunch (Fig. 1, $B_f \sim I_{b1}$). In both cases, the goal was to demonstrate uniform focusing of the bunch after precursor and possibility of reduction in energy spread due to the bunch head being in the deceleration phase and the bunch tail in the acceleration phase.

## RESULTS OF SIMULATION

Fig. 2 shows the case of the greatest achieved focusing of the bunch after precursor when lowest bunch radius in center is achieved (second bunch, movement from left to right). First consider the case (a) of Short Gaussian positron bunch case. (Figure 1(a)).

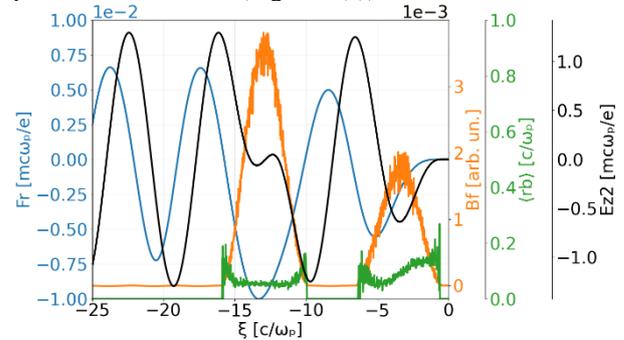

*Fig. 2. Radial component of the Lorentz force $F_r$ (blue) acting on bunch particles, azimuthal component of the magnetic field $B_f$ (orange), average bunch radius $<r_b>$ (green), off-axis longitudinal electric field $E_{z2}$ (black). For all quantities r=0. ξ=z-ct. Simulation time is 18 $\omega_{pe}^{-1}$. Short Gaussian bunch.*

In the case under consideration, the length of the precursor bunch and the second bunch is equal to the length of the plasma wave $\lambda_{pe}=2\pi c/\omega_{pe}$=10.56 cm, and the distance between precursor and second bunch is $\lambda_{pe}/2$ (Fig. 1-3). Figure 3 shows the dependences of the longitudinal field on the system axis and the averaged longitudinal electric field $<E_z>$.

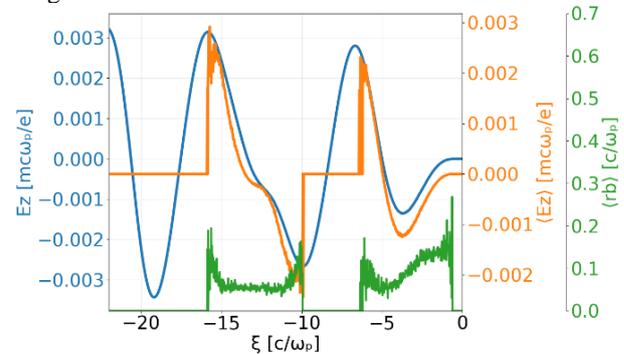

*Fig. 3. Longitudinal component of the electric field on the axis $E_z$ (blue), average longitudinal electric field at the bunch cross section $<E_z>$ (orange), average bunch radius $<r_b>$ (green). ξ=z-ct. t=18 $\omega_{pe}^{-1}$. Short Gaussian bunch.*

On the Fig. 3 $<E_z>(\xi) = (\Sigma_i [E_z(r_i, \xi) * q_{b,i}]) / (\Sigma_i [q_{b,i}])$, where $E_z(r_i, \xi)$ is the value of the longitudinal field at the location of the i-th particle; *i* is an index that runs through all particles of the bunch cross section; $q_{b,i}$ - charge of the i-th macroparticle; $<E_z>(\xi)$ is average field.

In Fig. 2 it is seen that the positron bunch-precursor excites the wakefield. The tail of the bunch- precursor is focused. The focus of the central part of the second positron bunch after bunch-precursor is clear.

Total on-axis longitudinal electric field $E_{z2}$ is 0. In addition, it is clear (Fig. 3) that the head and tail of the second short bunch are focused worse, which does not worsen the full case. Figure 3 also allows us to estimate the dynamics of the radii of the bunch-precursor and second positron bunches in the considered cases.

It is clear that in the central part of the second bunch (at a length of approximately 70% of the total bunch length) a "plateau" is observed, i.e. a uniform value $r_b$ equal to 0.05 $c/\omega_{pe}$ at t=18 $\omega_{pe}^{-1}$. This indicates the advantage of using such a configuration of plasma lens.

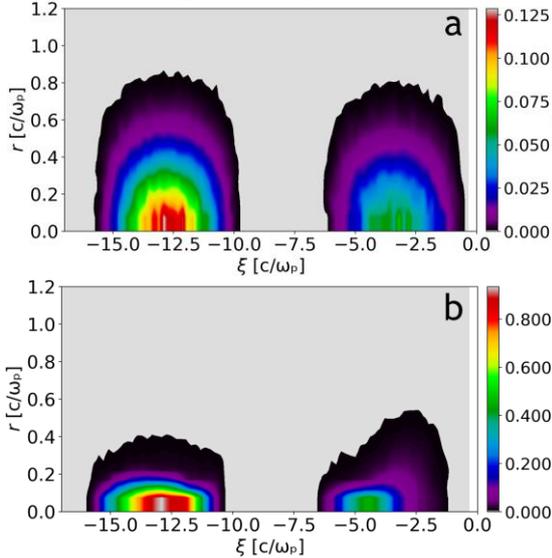

Fig. 4. Distribution of positron bunch density. Case 1) short gaussian bunch. (a) t=1 $\omega_{pe}^{-1}$, (b) t=18 $\omega_{pe}^{-1}$. The bunches move from left to right. For second short bunch $r_b/r_a=0.48$.

Figure 4 also allows us to estimate the dynamics of the second positron bunch. From Fig. 4(b) the focusing of the positron second bunch after precursor is obvious. By the time t=18 $\omega_{pe}^{-1}$ the bunch becomes focused and its radius decreases by 2.6 times. Figure 3 also shows that the average longitudinal electric field at the bunch cross section $<E_z>$: the acceleration of the tail of the bunch is compensated by the same deceleration of the head.

This will obviously contribute to the reduction of the energy spread in numerical simulations and experiments. Fig. 5, 6 show the simulation results for case 2) of a long second positron bunch after bunch-precursor. The head and tail of the bunch have Gaussian charge distributions with length $\lambda_{pe}/2$. The central part has a homogeneous charge (current) distribution. The length of the homogeneous central part is $\lambda_{pe}$ (Fig. 1(b)). An analysis of Figures 5 and 6 proves that the previously considered dynamics of case 1) for a short Gaussian bunch is generally valid in this case for a long bunch.

At the same time, there is an important difference. Due to the fact that the bunch is long, the focusing force acting on it quickly becomes non-uniform.

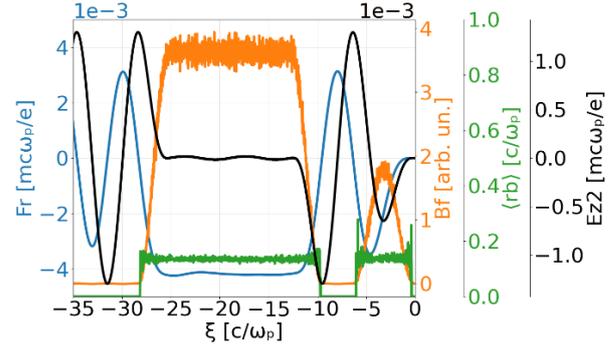

Fig. 5. Radial component of the Lorentz force $F_r$ (blue) acting on bunch particles, azimuthal component of the magnetic field $B_f$ (orange), average bunch radius $<r_b>$ (green), off-axis longitudinal electric field $E_{z2}$ (black). $\xi=z-ct$. $t=6$ $\omega_{pe}^{-1}$. Long bunch.

A uniform value of the focusing force is observed in the bunch area up to t=6 $\omega_{pe}^{-1}$. Therefore, it becomes necessary to use the "pulse focusing" mode.

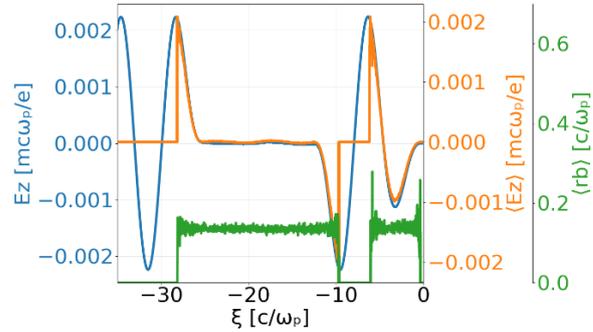

Fig. 6. Longitudinal component of the electric field on the axis $E_z$ (blue), average longitudinal electric field at the bunch cross section $<E_z>$ (orange), average bunch radius $<r_b>$ (green). $\xi=z-ct$. $t=6$ $\omega_{pe}^{-1}$. Long bunch.

After demonstrated the focusing of the second positron bunch after bunch-precursor, it is possible to add the sequence of the several positron bunches (after second bunch) with intervals of $0.5\lambda_{pe}$ between them. Then it can be obtained a sequence with the same value of the focusing force starting from second bunch. (Fig. 7).

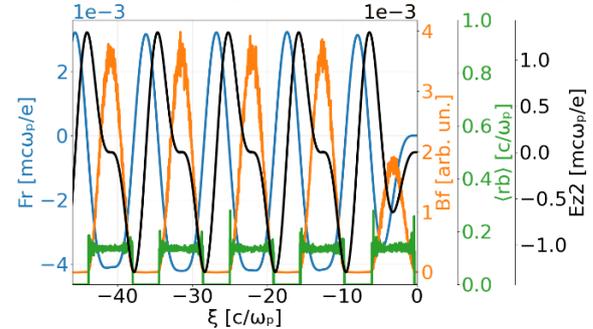

Fig. 7. Radial component of the Lorentz force $F_r$ (blue) acting on bunch particles, azimuthal component of the magnetic field $B_f$ (orange), average bunch radius $<r_b>$ (green), off-axis longitudinal electric field $E_{z2}$ (black). $\xi=z-ct$. Sequence of short positron bunches after first bunch-precursor.

Fig. 8(a) shows a focused second bunch after a precursor (Fig. 4(b)) with reflection about the axis of symmetry r=0. The distribution approaching Gaussian (semi-Gaussian with natural distortions) is maintained with time after focusing (Figure 8 (b, c)). This allows determine the radius of the positron bunch such way: the half-width limited by 3σ. Plots for several points are for illustrative purposes. In different parts of the bunch along ξ authors obtained semi-Gaussian distribution.

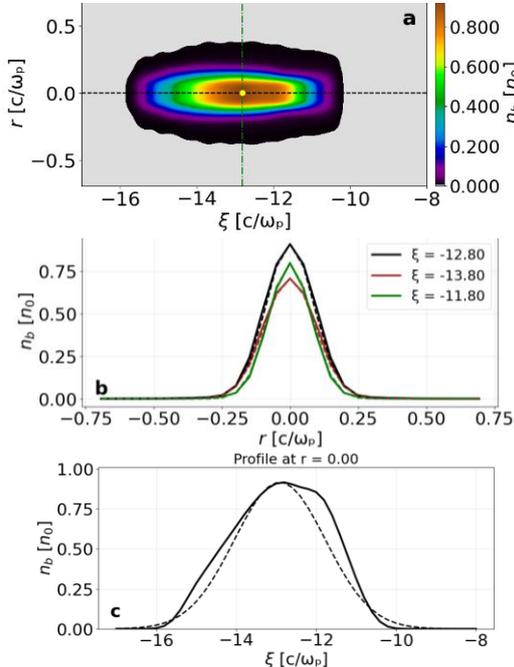

*Fig. 8. Focused second bunch after the precursor from Fig. 4(b). (a) symmetric reflection with respect to the r=0, the dot is the maximum density value $0.934n_0$; (b) slices at fixed ξ; (c) slices at fixed r=0. $t=18\ \omega_{pe}^{-1}$. The dashed line represents the Gaussian approximation.*

## CONCLUSIONS

This paper presents a method for the focusing and of positron bunches by using a plasma lens in linear regime.

Numerical simulations show that by positioning the bunch in the correct phase of the wakefield where its head is decelerated while its tail is accelerated, high-quality transverse focusing is achieved, reducing the bunch radius by a factor of 2.6.

The possibility of focusing of positron bunches sequences with identical and uniform focusing is demonstrated.

## ACKNOWLEDGMENTS

The authors are grateful to Prof. Wim Leemans (DESY) for productive scientific discussions.

## ПЛАЗМОВА ЛІНЗА ДЛЯ ФОКУСУВАННЯ ПОЗИТРОННИХ ЗГУСТКІВ

*Д. С. Бондар, К. А. Ліндстрем, В. І. Маслов, І. М. Оніщенко*

Розробка ефективних плазмових лінз для фокусування позитронних згустків у перспективних прискорювачах залишається значним викликом, оскільки нелінійні режими не здатні створювати ефективних фокусуючих каналів для позитронів. У цій роботі представлено метод фокусування та покращення якості позитронних згустків з використанням плазмової лінзи, що працює в лінійному режимі. За допомогою числового моделювання досліджено два різні профілі позитронного згустка, що фокусується: чисто гаусів згусток та видовжений згусток із плоскою вершиною та Гаусовими фронтами наростання та спаду. Для обох конфігурацій результати демонструють можливість досягнення якісного поперечного фокусування. Крім того, окрім фокусування, запропонована система потенційно дає можливість зменшити розкид енергії позитронних згустків послідовності після згустка-попередника.